# Manipulating Weyl quasiparticles by orbital-selective photoexcitation in WTe$_2$

Meng-Xue Guan,[1,2,4] En Wang,[1,2,4] Pei-Wei You,[1,2] Jia-Tao Sun,[1] and Sheng Meng[1,2,3,*]

[1]*Beijing National Laboratory for Condensed Matter Physics and Institute of Physics, Chinese Academy of Sciences, Beijing 100190, China.*

[2] *School of Physical Sciences, University of Chinese Academy of Sciences, Beijing 100190, China.*

[3]*Songshan Lake Materials Laboratory, Dongguan, Guangdong 523808, China.*

[4] These authors contributed equally: Mengxue Guan, En Wang.

*email: smeng@iphy.ac.cn

**Optical control of structural and electronic properties of Weyl semimetals allows development of switchable and dissipationless topological devices at the ultrafast scale[1-9]. An unexpected orbitial-selective photoexcitation in type-II Weyl material WTe$_2$ is reported under linearly polarized light (LPL), inducing striking transitions among several topologically-distinct phases. The symmetry features of atomic orbitals comprising the Weyl bands result in asymmetric electronic transitions near the Weyl points, and in turn an anisotropic response of interlayer shear motion with respect to linear light polarization, when a near-infrared laser pulse is applied. Consequently, not only annihilation of Weyl quasiparticle pairs, but also increasing separation of Weyl points can be achieved, complementing existing experimental observations. Our results provide a new perspective on manipulating the singularity of Weyl nodes and coherent control of electron and lattice quantum dynamics simultaneously.**

Three-dimensional Weyl semimetals (WSMs) are novel topological phases of matter, in which the chiral Weyl node can be viewed as pseudo-magnetic monopoles in momentum space and the magnetic charge is determined by chirality[1-7]. These magnetic monopoles have direct effects on the motion of electrons, providing an ideal platform to explore the nonlinear optoelectronic responses that related to topology in gapless materials[8,9]. The nonlinear responses of the WSMs, such as the second harmonic generation[10], nonlinear Hall effect[11-14] and topological phase transition[15-17], are of great importance in probing the fundamental properties of quantum materials as well as for applications in optoelectronic devices and solar cells.

Although a nonlinear optical process that involves high-energy excited states may not directly capture the low-energy singularity at the Weyl nodes, the ground state and low-energy excited states are characterized by the geometry and topological nature of the Bloch wavefunctions at the node points. For example, enhanced bulk photovoltaic effect (BPVE) are reported in two typical WSMs, i.e., type-I WSM TaAs[18] and type-II WSM $TaIrTe_4$[19], by mid-infrared light illumination. The BPVE is a nonlinear optical response that intrinsically converts linearly polarized light into electrical dc current through the shift of charge centre during the interband excitation[20-23]. The substantially enhancement of the generated photocurrent, i.e., shift current, is ascribed to the Berry flux field singularity of the Weyl nodes. The results indicate that the carrier excitation around the Weyl nodes display a range of unique behaviors, which are highly sensitive to the laser polarization, helicity and wavelength.

The connection between macroscopic nonlinear optoelectronic responses and band topology of the WSMs are routinely established by models based on ground-state calculations and symmetry analysis[24,25]. However, there are two limitations: i) The lack of the real-time dynamics of excited electrons in momentum-space and the corresponding real-space; ii) The electron-lattice and photon-lattice coupling are intrinsically ignored, which are of crucial importance in inducing topological phase transitions. Real-time time-dependent density functional theory molecular dynamics (TDDFT-MD) is a technique that may overcome these difficulties, and has been applied to unravel the interplay among different degrees of freedom in materials under photoexcitation[26-30].

In this work, we demonstrate that the photocurrent can be selectively excited in bulk WTe$_2$, a type-II WSM, based on ab initio TDDFT-MD quantum simulations[31-33]. The current direction depends on the laser polarization as well as the photon energy, both of which play a vital role in determining the features of final excited states. The carriers can be excited around the Weyl node when a near-infrared laser pulse (photon energy $\hbar\omega = 0.5{\sim}0.8$ eV) is applied. Asymmetric excitation at the space-inversed $\boldsymbol{k}$-points around the Weyl node is achieved, depending on linear light polarization and atomic orbital features of Weyl bands. In contrast, higher energy photons would excite electrons to high-energy bands, making the process polarization isotropic and of marginal relevance to the Weyl physics. The anistropic photocurrent drives distinctive interlayer shear displacements, providing an ultrafast way for modulating topological properties, e.g., annihilation or increasing separation of Weyl nodes.

**Selective excitations under linearly polarized light**

The experimental geometry of the bulk WTe$_2$ is adopted, which is characterized by an orthorhombic ($T_d$) unit cell without inversion symmetry[34]. The covalently bonded W atoms form a zigzag W-W chain along $a$-axis, leading to distinct anisotropy in the two dimensional plane[35,36] (Fig. 1a). There are 8 Weyl points (WPs) in the $k_z = 0$ plane, and two of the WPs are shown in Fig. 2a, where the valence and conduction bands constituting the Weyl nodes are labeled as band-0 and band-1, respectively. Here, 'WP1' carries a Chern number $C = +1$ and locates nearly on the high-symmetry line Γ-X and is labeled as 'W' in Fig. 2b.

To study the optoelectronic responses of $T_d$-WTe$_2$, linearly polarized laser beams with time-dependent electric field $E(t) = E_0 cos(\omega t) exp[-(t-t_0)^2/2\sigma^2]$ are applied. The photon energy, width, and intensity are set as 0.6 eV, 10 fs and 6.5×10$^{10}$ W/cm$^2$, respectively (Supplementary Fig. 1a). This setup allows us to reproduce a laser fluence (ca. 0.6 mJ/cm$^2$) similar to experimental measurements[37]. There are 0.62% and 0.18% of valence electrons excited to the conduction bands after laser pulse ends ($t = 100$ fs), when the laser polarization is along the crystallographic *a*-axis (LP-a) and *b*-axis (LP-b), respectively (Supplementary Fig. 1b) .

Figure 3a and 3b show the energy distribution of excited electrons and holes after the laser pulse. It is clear that almost all electrons are promoted to low-energy excited states in the range of $E - E_f < 1$ eV ($E_f$ is the Fermi level) in both cases. Furthermore, a major part of electrons are promoted to states whose energy are less than 0.3 eV above $E_f$, the grey shaded region in Fig. 2b, indicating that these excitations involve transitions between the bands forming or nearing the Weyl cone. We find that there are two kinds of inter-band transitions: most of electrons are promoted to band-1 under LP-a excitation (Fig. 3c), while band-2 (band-2 is the adjacent, higher-energy conduction band) is dominant under LP-b excitation (Fig. 3d) Note that the tilted Weyl node leads to the partial occupation of band-1 along the W-X line in the Brillouin zone (BZ), where the ***k***-points energies are below the $E_f$. The carrier transitions from the valence bands to the band-1 are thus forbidden along this direction due to the Pauli blocking effect. Electronic excitations take place at ***k***-points along W-Γ (W-X) direction when the laser polarization is along *a*-axis (*b*-axis) (Fig. 2c). Therefore, optical transitions around the Weyl node can be controlled by manipulating laser polarization.

**Unique orbital features around the Weyl nodes**

The polarization controlled selective excitation arises from the asymmetric distribution of electron orbitals around the Weyl node. By projecting electronic wavefunctions of these bands onto atomic orbitals, one can clearly see that the dominant orbital components of band-0 and band-1 are exchanged due to the band inversion around the WP1 (Fig. 2b). According to the optical selection rule, the transition probability depends on the geometry of the orbitals, i.e., $M_{ij} = \langle \psi_j | \hat{r} | \psi_i \rangle$, where $M_{ij}$ is the transition matrix element between the initial state $|\psi_i\rangle$ and final state $|\psi_j\rangle$. We find that $\langle d_{z^2} | \hat{x} | p_x \rangle$ and $\langle d_{z^2} | \hat{x} | d_{xz} \rangle$ are two dominant transition pathways along W-Γ under LP-a excitation, whereas $\langle d_{z^2}|\hat{y}|d_{yz}\rangle$ is the most important pathway along W-X under LP-b excitation (Note, crystallographic *a*-axis and *b*-axis correspond to the real-space *x*-axis and *y*-axis, respectively.) (Fig. 2c). These characteristic transitions are forbidden by symmetry under the other excitation condition, i.e., $\langle d_{z^2}|\hat{y}|p_x\rangle = \langle d_{z^2}|\hat{y}|d_{xz}\rangle = \langle d_{z^2}|\hat{x}|d_{yz}\rangle = 0$. Therefore, in addition to regular spin selection rules around the Weyl points, the unique wavefunction features lead to a new kind of linear-polarization dependence of photoexcitation.

We then discuss the similarities and discrepancies between the above selective excitation and the well-known chirality selection rule[38,39]. Both of them determine the carrier transition pathways near the Weyl cone. In the latter case, the absorption of a circularly polarized photon flips the spin, resulting in asymmetric excitations along the driving direction. For example, for a $\chi = +1$ Weyl fermion and a right-handed circularly polarized (RCP) light, the transition is allowed on the +*k* side but forbidden on the −*k* side of the Weyl node due to the conservation of angular momentum (Fig. 1c). The LP-a light in the present work, plays a similar role in introducing transitions as what RCP light do to ideal Weyl points (Fig. 1d). However, the selective excitation reported here is mutually determined by the linear polarization and atomic orbital features due to the combination of inversion symmetry breaking and finite tilts of the Weyl dispersion, whereas in chirality selection rule, the coupling between the laser helicity and the chirality of the Weyl node is the dominant factor.

**Photocurrent from photoexcited Weyl semimetal**

The asymmetric carrier excitations at the Weyl points will contribute to significant photocurrent generation[39,40]. The induced photocurrent along *a*-axis ($I_a$) and *b*-axis ($I_b$) under LP-a and LP-b excitations are shown in Fig. 4a,b. Compared to the currents that are parallel to the incident polarization, the magnitudes of the perpendicular components are much smaller and show different oscillation behaviors. A nearly unidirectional current along the *b*-axis emerges with LP-a excitation, thanks to orbital symmetries and corresponding shift current in $T_d$-WTe$_2$. Detailed analysis of the microscopic mechanisms warrants further theoretical and experimental explorations, which is beyond the scope of this work.

The time integrals of $I_a$ and $I_b$ are shown in Fig. 4c,d, implying the accumulative effect of carrier excitation. After the laser pulse ends, the two polarizations along *a*-axis and *b*-axis induce net charges with the opposite directions, whose real-space distribution can be described by the charge density difference (CDD) (Fig. 4e,f). Here, time-dependent CDD is shown on a plane cutting along the top layer of the unit cell and is defined as $\Delta\rho(\boldsymbol{r},t) = \rho(\boldsymbol{r},t) - \rho(\boldsymbol{r},0)$, where $\rho(\boldsymbol{r},t)$ is the density at time *t*. The inhomogeneous carrier distribution will introduce a built-in electric field in the material and couple with specific phonon modes, providing possibilities to induce phase transitions.

**Anisotropic response of interlayer shear motion**

Ultrafast symmetry switches were recently observed in $T_d$-WTe$_2$ and MoTe$_2$, both companying with interlayer shear displacements along the *b*-axis[37,41]. Sie *et al.* demonstrated that the shear motion always starts from the $T_d$ phase to the centrosymmetric (1T′ or 1T′(*)) phase regardless of polarization under the terahertz fields[37]. Here, we show that when the photon energy is specifically tuned, strongly anisotropic photoexcitaion will be achieved.

Atomic movements are launched by the electron-phonon couplings, whose direction is determined by the elaborate electronic transitions. Selective excitation of WSMs emerge only if the excitation involves transitions between bands forming (or nearing) the Weyl cone, leading to polarization-dependent shear mode. However, when a terahertz laser pulse is applied, the photon energy (e.g., 23 THz ≈ 0.095 eV) is too low to promote electrons to higher-lying conduction bands (e.g., band-2), therefore, only transitions between band-1 and band-0 (path 'LP-a' in Fig. 2c) is allowed regardless laser polarization. On the other hand, a much higher photon energy would excite electrons to bands far above the Weyl points, making the process marginal relevance to Weyl physics (path ② in Fig. 2b). Therefore, polarization-isotropic interlayer displacement is expected in the above two excitation conditions (i.e., THz and visible excitations).

To describe the interlayer motion, averaged atomic displacement is calculated,

$$\Delta y(t) = \frac{1}{N}\sum_{i=1}^{N}\{y_i(t) - y_i(0)\}.$$

The summation runs over all atoms in the top (or bottom) layer in the unit cell, and $y_i(t)$ is the time-dependent position of atom $i$ along the $b$-axis. In Supplementary Fig. 2, we show that when the photon energy is 1.5 eV, the interlayer movement always starts from the $T_d$ phase to the centrosymmetric phase, which is consistent with experimental observations[37,41]. Similar methods have also been adopted to describe the interlayer movement along $a$-axis and $c$-axis, however, negligible displacements are observed (Supplementary Fig. 3). It is ascribed to the strongest electron-phonon coupling along $b$-axis due to the existence of effective interlayer-shearing phonon mode[42,43].

Furthermore, we predict that the polarization-anisotropic response of interlayer shear displacement is achieved when a near-infrared laser pulse (ca. 0.6 eV) is applied, as shown in Fig. 5a,b. It is clear that the adjacent layers move with nearly identical velocity ($\approx 0.25$ Å/ps), but along opposite directions. With LP-a excitation, the bottom layer move towards the positive direction of the $b$-axis, while the top layer move along the negative direction, leading to the restoring of inversion symmetry, in line with the experimental measurement[37]. However, the relative motion between the two layers is reversed under LP-b excitation, and the non-centrosymmetric order is further enhanced. It implies that the shear motion can be controlled towards either a centrosymmetric trivial phase or a non-centrosymmetric topological phase. The polarization-anisotropic is related to the singularity of the Weyl nodes, and the suitable photon energy is in the range of 0.5~0.8 eV (Supplementary Fig. 4).

To compare our simulation results with experimental data, we define non-centrosymmetric degree (NCD) of WTe$_2$, which can be verified using time-resolved second-harmonic generation (SHG) technique[37] or by monitoring the interlayer shear displacements (Supplementary Fig. 5). Figure 5c shows the normalized NCD under LP-a excitation, for the initial $T_d$ phase, NCD is 1; for centrosymmetric phase, NCD is 0. The time-evolution of NCD match well with the experimentally detected SHG signal, justifying the reliability of the present approach. Limited by the unprecedented computational cost of first-principles excited-state dynamic simulations, the maximum time we could simulate is less than 1 ps, which describes the early stage of the photoinduced phase transition. In Supplementary Fig. 5(c), we show that after $t = 600$ fs, the NCD of $WTe_2$ tends to return to its initial state under both LP-a and LP-b excitations. Based on that, we expect the interlayer displacements might exhibit an oscillation behavior in later stages.

**Light-controlled Weyl nodes**

The polarization-dependent shear motion provides an effective way to control topological properties of $WTe_2$. Depending on the direction of interlayer shearing, all WPs of opposite chirality will be annihilated under LP-a excitations, or separated further to have extremely long Fermi arcs under LP-b excitations. Figure 5d show the time-dependent positions of WPs, where WP1 and WP2 are two nearest WPs with the Chern number +1 and -1, respectively. At equilibrium ($t = 0$ fs), the WPs are separated by 0.7% of the reciprocal lattice vector $|\boldsymbol{G}_2|$. Under LP-a excitation, the WPs move towards each other and the annihilation occurs when $t = 150$ fs. Under LP-b excitation, the WPs are pushed away from each other in momentum space, leading to more robust, ideally separated WPs, e.g., the WPs are separated by 5% of $|\boldsymbol{G}_2|$ at $t = 300$ fs. Thus a large separation is crucial in realizing a giant quantum anomalous Hall effect, with Hall conductivity is proportional to the separation between Weyl nodes[13,44,45]. Further increasing the displacement, WP2 annihilates with its mirror image of opposite chirality ($t = 600$ fs), a Lifshitz transition occurs from a topological semimetal with eight WPs to one with four WPs. Therefore, the number of the WPs can be modulated freely to zero, four and eight by controlling the photon energy and the polarization of the laser pulse (Fig. 5e). To show the evolution of WPs separation, the phase diagram as functions of photon energy and incident direction is constructed, as shown in Fig. 1b.

**Discussion**

Our *ab initio* TDDFT-MD simulations reveal that the phase transition in type-II WSM WTe$_2$ can be controlled by orbital-selective electronic excitations, mediated by effective electron-phonon coupling. Polarization-anisotropic interlayer displacement is predicted when the carrier transitions are in the vicinity of the Weyl cone, where the orbital-dependent excitations of WSMs emerge with a suitable photon energy. In this scenario, the topological phase transitions can be controlled towards either annihilating all WPs or inducing largely separated WPs. However, polarization-isotropic response of shear motion is expected if the transitions are far away from the Weyl cone, inducing only WPs annihilation. Our work provides a new insight into controlling Berry flux field singularity around the Weyl nodes, and the *ab initio* approach adopted here might be useful for understanding a wide range of non-linear responses of topological materials.

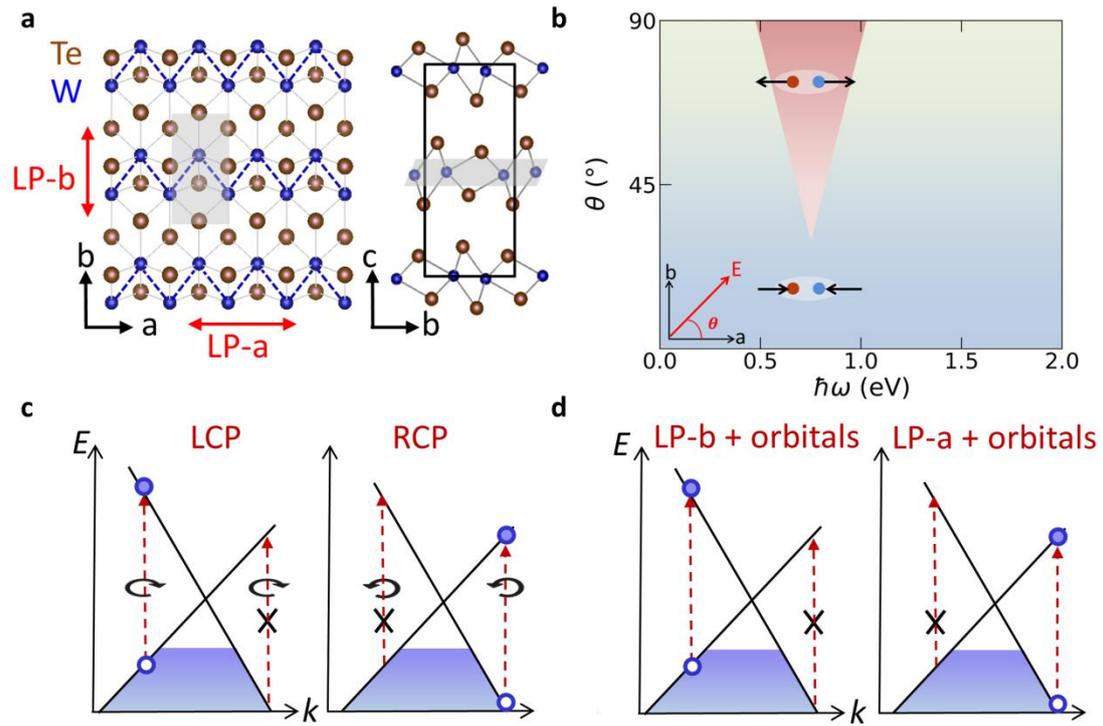

**Fig. 1| Illustration of selective photoexcitation in WTe$_2$. a,** Lattice structure of $T_d$-WTe$_2$: top view (*a-b* plane) and side view (*b-c* plane), the dashed lines indicate the W-W zigzag chain along *a*-axis. The top layer is denoted by the grey shaded plane. **b,** Phase diagram of laser-driven $T_d$-WTe$_2$ topological phase transition on the dependence of photon energy $\hbar\omega$ and incident angle $\theta$. The linearly polarized laser pulse propagate on the *a-b* plane with the propagation direction as $\theta$. **c,** Schematics of the chiral selection rule of the $\chi = +1$ Weyl node in momentum space. **d,** Schematics of the orbital-dependent selective excitations when the laser pulses are linearly polarized along the *a*-axis (LP-a excitation) and *b*-axis (LP-b excitation).

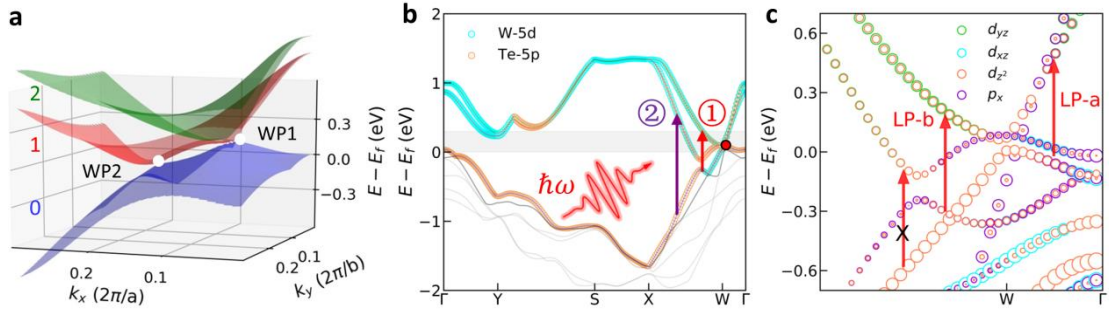

**Fig. 2| Band structure of WTe$_2$. a,** Band structure of $T_d$-WTe$_2$ in the vicinity of two Weyl nodes. **b,** Band structure along the high-symmetry lines in the Brillouin zone. The radiuses of cyan and yellow dots indicate the weight of W-5$d$ and Te-5$p$ orbitals. The red dot represents the position of WP1. Arrows ① and ② show the carrier transitions that near or far away from the Weyl node. **c,** The magnification of the band structure along X-Γ, and with more detailed atomic orbital information. The red arrows represent the transitions with the laser linearly polarized along the $a$-axis (LP-a excitation) and $b$-axis (LP-b excitation) with photon energy of 0.6 eV.

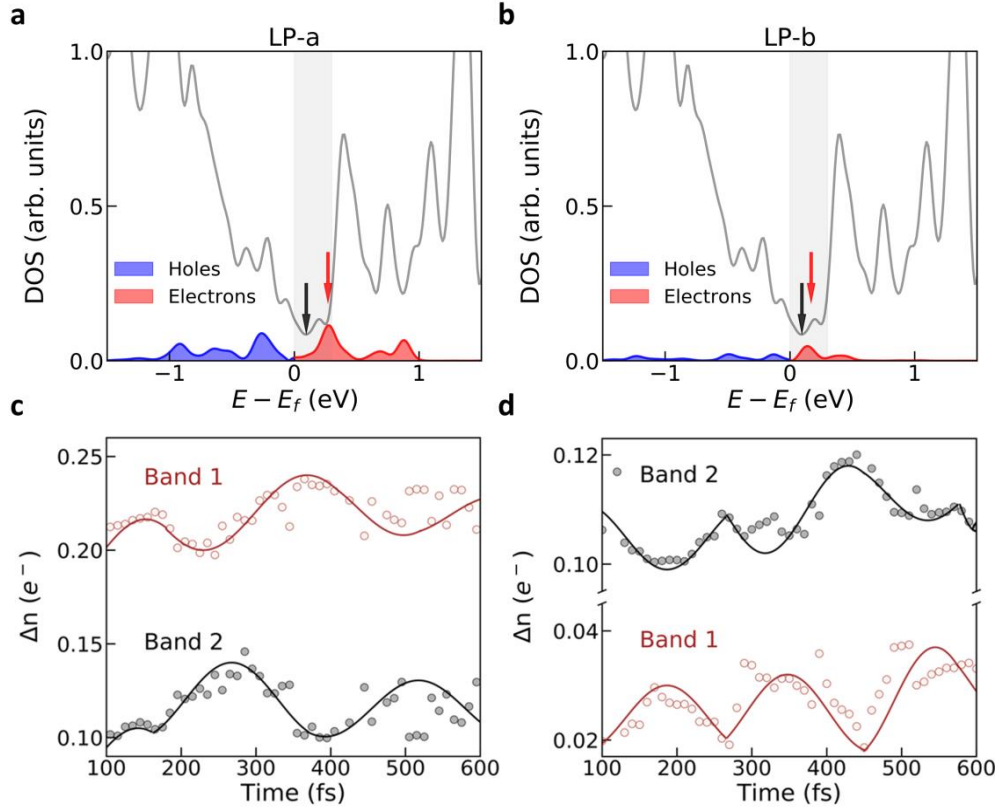

**Fig. 3| Carrier excitations in the momentum space. a,** Energy distribution of the excited electrons and holes at $\hbar\omega = 0.6$ eV excitation linearly polarized along the *a*-axis (LP-a). The grey shaded area indicates the low-energy excitation region that near the Weyl node. The red and black arrows indicate the peak of the electronic excitation and the energy position of WP1, respectively. **c,** Time evolution of the electronic occupation of band-1 (brown dots) and band-2 (grey dots) under LP-a excitation. The colored lines are guides to the eyes based on the oscillations of the electronic occupation. **b**, **d,** are analogous to **a** and **c**, but are the results when laser linearly polarized along the *b*-axis (LP-b).

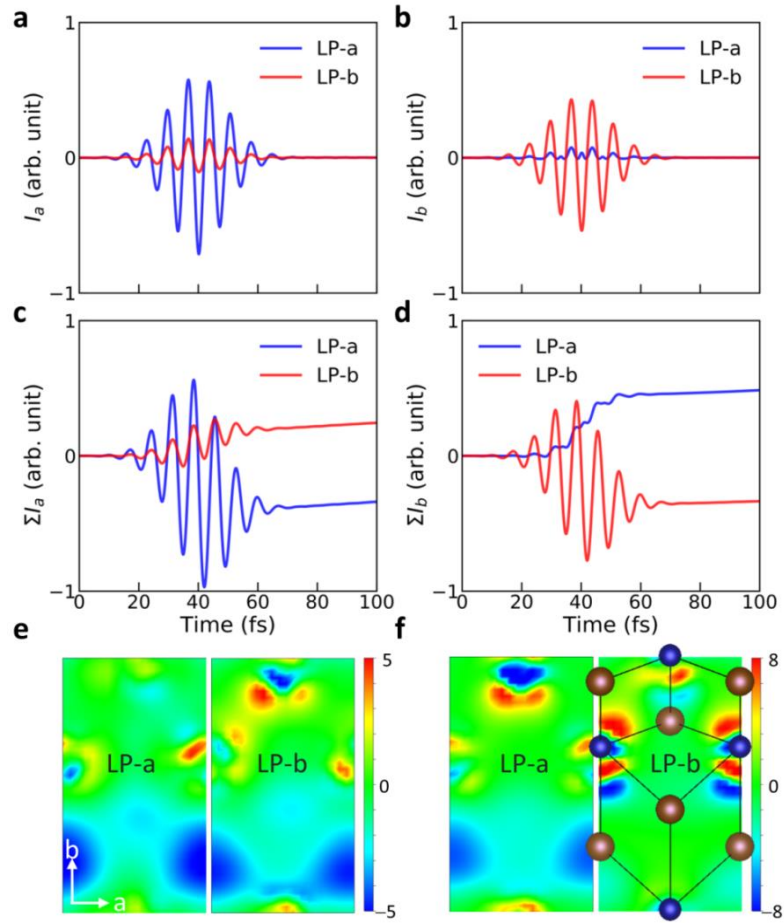

**Fig. 4| Laser polarization dependent photocurrent. a, b,** Laser induced current along $a$-axis ($I_a$) and $b$-axis ($I_b$) under LP-a and LP-b excitations. **c, d,** Time integrals of $I_a$ and $I_b$. **e, f,** Charge density differences (CCD) between ground and excitation states at 50 fs (**e**) and 150 fs (**f**). The unit of CDD is $10^{-3} \times e/a_0^3$, $a_0$ is the Bohr radius. The blue (red) region represents the increase (decrease) in charge density.

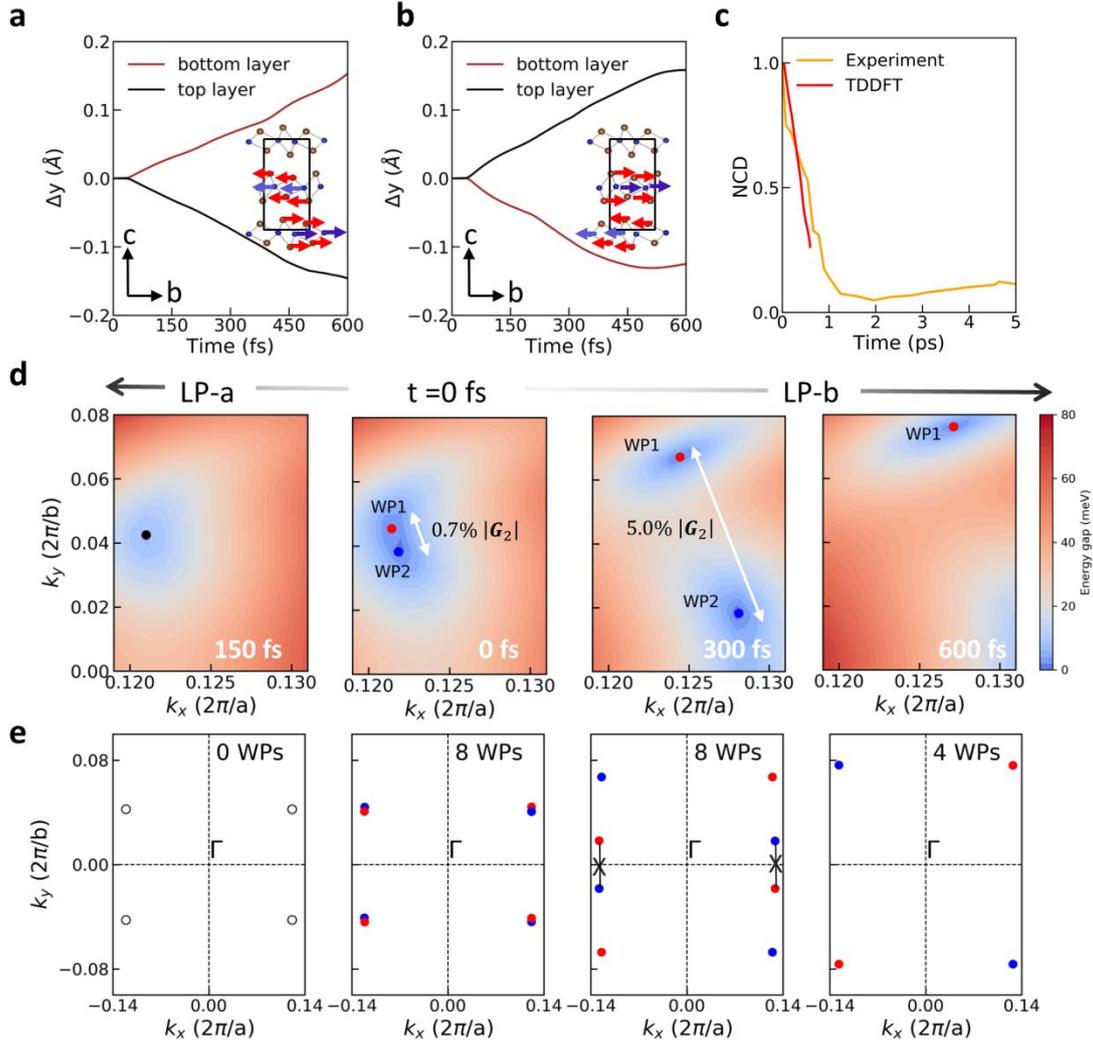

**Fig. 5| Polarization-anisotropic response of interlayer shear displacement and the induced topological phase transitions. a,** Interlayer shear displacement under LP-a excitations. **b,** Interlayer shear displacement under LP-b excitations. The inset in each panel shows the corresponding shear mode. **c**, Comparison of non-centrosymmetric degree (NCD) between TDDFT result under LP-a excitation and experimental detected SHG signals taken from Ref. [37]. The SHG time trace is measured at pump field strength of 10 MV cm$^{-1}$, and the laser pulse wavelength is 2.1 μm. **d,** The two nearest WPs in the $k_z = 0$ plane (WP1, red; WP2, blue) at various shear displacement conditions. **e,** Schematic illustration of the number of WPs and their separations in the $k_z = 0$ plane.

## Methods

The TDDFT-MD calculations are performed using the time dependent ab initio package (TDAP) as implemented in SIESTA[31-33]. The bulk $WTe_2$ in its $T_d$ phase is simulated with a unit cell of 12 atoms with periodical boundary conditions. Numerical atomic orbitals with double zeta polarization (DZP) are employed as the basis set. The electron-nuclear interactions are described by Troullier-Martins pseudopotentials, PBE functional[46]. An auxiliary real-space grid equivalent to a plane-wave cutoff of 250 Ry is adopted. To make a good balance between the calculation precision and cost, a Γ-centered $6 \times 5 \times 3$ *k*-point grid is used to sample the Brillouin zone. The coupling between atomic and electronic motions is governed by the Ehrenfest approximation[47]. During dynamic simulations the evolving time step is set to 0.05 fs for both electrons and ions in a micro-canonical ensemble.

DFT calculations of the electronic band structure of $WTe_2$ at different interlayer displacement Δy (Fig.5c and 5d) are performed by Vienna ab initio simulation package (VASP 5.4)[48]. Exchange-correlation effects are treated at the level of the generalized gradient approximation (GGA) through the PBE functional. The Projector-augmented wave (PAW) potentials with valence electronic configurations of {6*s*2, 5*d*4} for W and {5*s*2, 5*p*4} for Te are employed in conjunction with a plane-wave energy cutoff parameter of 300 eV. For self-consistent electron density convergence, a Γ-centered $12 \times 10 \times 6$ *k*-point grid and Gaussian smearing with a smearing parameter of 0.05 eV are used. To obtain the positions of the Weyl Points (WPs) during the ultrafast topological phase transition, Wannier90[49,50] interface and WannierTools[51] are used to investigate the topological properties.

## Data availability

The data that support the plots within this paper and other findings of this study are available from the corresponding author upon reasonable request.

## Acknowledgements

We acknowledge partial financial support from the National Key Research and Development Program of China (No. 2016YFA0300902, 2015CB921001 and 2016YFA0202300), National Natural Science Foundation of China (No. 91850120, 11774396 and 11934003), and "Strategic Priority Research Program (B)" of Chinese Academy of Sciences (Grant No. XDB07030100 and XDB30000000). The authors acknowledge helpful discussions with Prof. Shengbai Zhang.

## Author contributions

S.M. designed the research. Most of the calculations were performed by M.G. and E.W. with contributions from all authors. All authors contributed to the analysis and discussion of the data and the writing of the manuscript. M.G. and E.W. contributed equally to this work.

## Competing interests

The authors declare no competing interests.

Supplementary Information for

# Manipulating Weyl quasiparticles by orbital-selective photoexcitation in WTe$_2$

Meng-Xue Guan,[1,2,4] En Wang,[1,2,4] Pei-Wei You,[1,2] Jia-Tao Sun[1], and Sheng Meng[1,2,3]

[1]*Beijing National Laboratory for Condensed Matter Physics and Institute of Physics, Chinese Academy of Sciences, Beijing 100190, China.*

[2] *School of Physical Sciences, University of Chinese Academy of Sciences, Beijing 100190, China.*

[3]*Songshan Lake Materials Laboratory, Dongguan, Guangdong 523808, China.*

**Supplementary Note 1: Applied laser waveform and the number of excited electrons**

The laser electric field $E(t)$ is described to be a Gaussian-envelope function,

$$E(t) = E_0\, cos(\omega t)\, exp\left[-\frac{(t-t_0)^2}{2\sigma^2}\right]. \qquad (S1)$$

Here, the width $\sigma$ is 10 fs, and photon energy $\omega$ is 0.6 eV. The laser field reaches the maximum strength $E_0 = 0.028$ V Å$^{-1}$ at time $t_0 = 40$ fs (Fig. S1a). Velocity gauge is used where the vector and scalar potential of the field $E(t)$ are $\vec{A}(t) = -c\int_0^t \vec{E}(t')dt'$ and $\Phi = 0$. Time evolution of the wavefunctions are then computed by propagating the Kohn-Sham equations in atomic units (a.u.),

$$i\frac{\partial}{\partial t}\psi_i(\vec{r},t) = \left[\frac{1}{2m}\left(\vec{p}-\frac{e}{c}\vec{A}\right)^2 + V(\vec{r},t)\right]\psi_i(\vec{r},t). \qquad (S2)$$

Time-dependent current can be obtained as,

$$J(t) = \frac{1}{2i}\int_\Omega d\vec{r} \sum_i \{\psi_i^*(\vec{r},t)\nabla\psi_i(\vec{r},t) - \psi_i(\vec{r},t)\nabla\psi_i^*(\vec{r},t)\}. \qquad (S3)$$

The dynamics of the excited electrons (Fig. S1b) are performed by projecting the time-evolved wavefuncitons ($|\psi_{n,\mathbf{k}}(t)\rangle$) on the basis of the ground-state Kohn-Sham

orbitals ($|\varphi_{n',\mathbf{k}}\rangle$)

$$\Delta n_e(t) = \frac{1}{N_\mathbf{k}} \sum_{n,n'}^{CB} \sum_{\mathbf{k}}^{BZ} |\langle \psi_{n,\mathbf{k}}(t)|\varphi_{n',\mathbf{k}}\rangle|^2, \tag{S4}$$

where $N_\mathbf{k}$ is the total number of the $\mathbf{k}$-points used to sample the BZ. All conduction-band electrons are summed up and $n$ and $n'$ are band indices. The $\eta$ is used to denote the percentage of valence electrons that are pumped to specified unoccupied bands. For instance, $\eta = 1\%$ means 1% of total valence electrons are pumped to conduction bands.

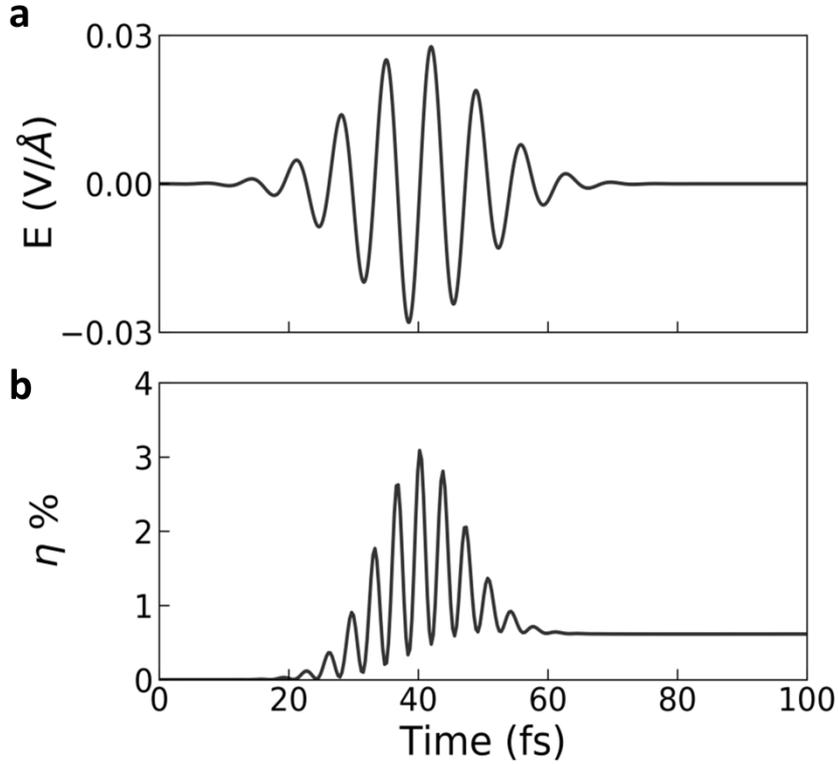

**Fig. S1| Laser waveform and the excited carriers. a,** Applied electric field along the in-plane direction of $T_d$-WTe$_2$ with the laser strength $E_0 = 0.028$ V/Å. The pulse width is 10 fs and photon energy is 0.6 eV. **b,** Number of excited electrons ($\eta$) upon photoexcitation from the valence bands to conduction bands with LP-a excitation. After the laser pulse ($t = 100$ fs), $\eta = 0.62\%$. For LP-b excitation, the fluctuation of excited electrons follows similar tendency as what we show here, but with $\eta = 0.18\%$.

**Supplementary Note 2: Electronic excitation and ionic movement with the photon energy of 1.5 eV**

Here, we show that when the carrier excitations are far away from the Weyl cone, the shear motion always starts from the $T_d$ phase to the centrosymmetric phase regardless of polarization, i.e., polarization isotropic. To demonstrate that, laser pulses with a photon energy of 1.5 eV are applied to $T_d$-WTe$_2$ (Fig. S2a). Following the same analysis methods in the main text, we find that both electrons and holes are excited to the energy levels far away from $E_f$ no matter what polarization is adopted (Fig. S2c,e). The resultant shear mode is polarization isotropic and towards restoring the inversion symmetry (Fig. S2d,f), consistent with previous experimental observations.

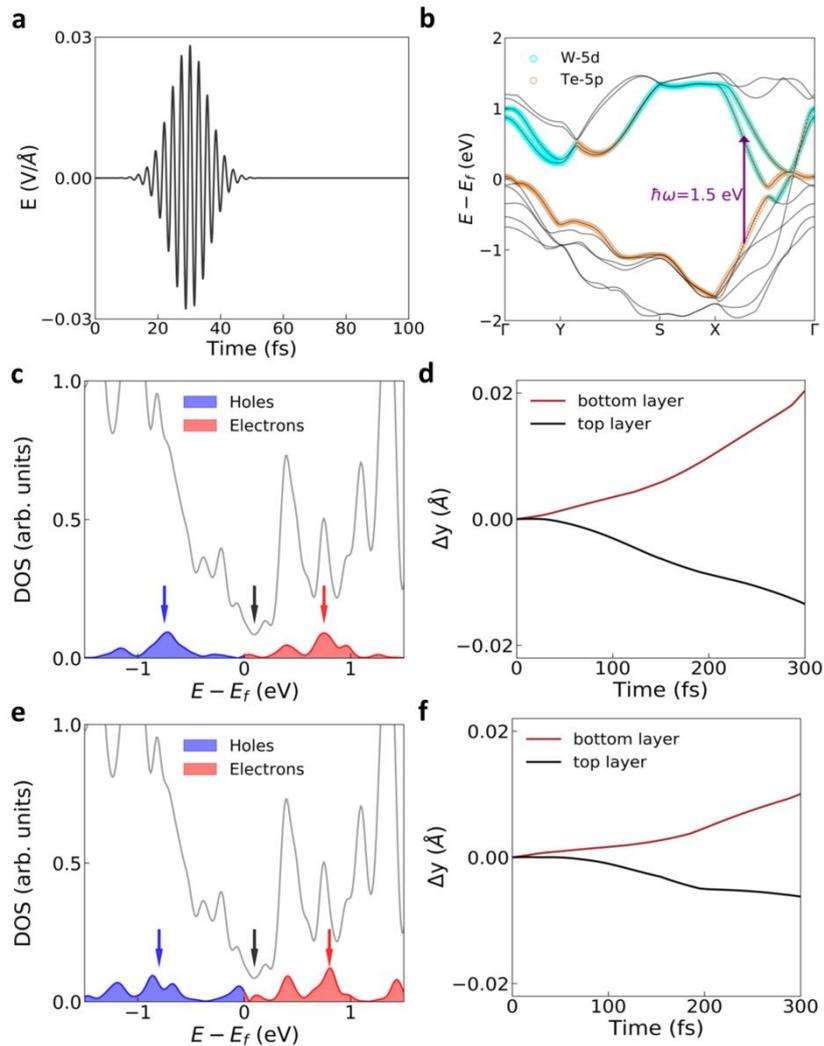

**Fig. S2| Carrier excitation and ionic movement with the photon energy of 1.5 eV.**

**a,** Applied laser pulse along the in-plane direction of $T_d$-WTe$_2$ with the laser strength $E_0 = 0.028$ V/Å and the photon energy is 1.5 eV. The pulse width is 6 fs and reaches it maximum at 30 fs. **b,** Schematic illustration of the carrier excitation under the light field. **c,** Energy distribution of the excited carriers with LP-a excitation, where the red and blue arrows denote the peaks of the excited electrons and holes, which locate at the energy level of 0.8 eV and −0.7 eV, respectively. **d,** Shear displacements of two layers along *b*-axis. **e, f,** are analogous to **c, d,** but under LP-b excitation.

**Supplementary Note 3: Interlayer motion along *a*-axis and *c*-axis with the photon energy of 0.6 eV**

The interlayer displacements along *a*-axis and *c*-axis are monitored as well and can be described as

$$\Delta x(t) = \frac{1}{N} \sum_{i=1}^{N} \{x_i(t) - x_i(0)\}, \quad \quad (S5)$$

$$\Delta z(t) = \frac{1}{N} \sum_{i=1}^{N} \{z_i(t) - z_i(0)\}, \quad \quad (S6)$$

where $x_i(t)$ and $z_i(t)$ are the time-dependent positions of atom *i* along the *a*-axis and *c*-axis, respectively. Comparing with the obviously movement along *b*-axis (Fig. 5a,b in the main text), negligible displacements are observed along these two directions (Fig. S3), indicating that the shear motion is only along the *b*-axis.

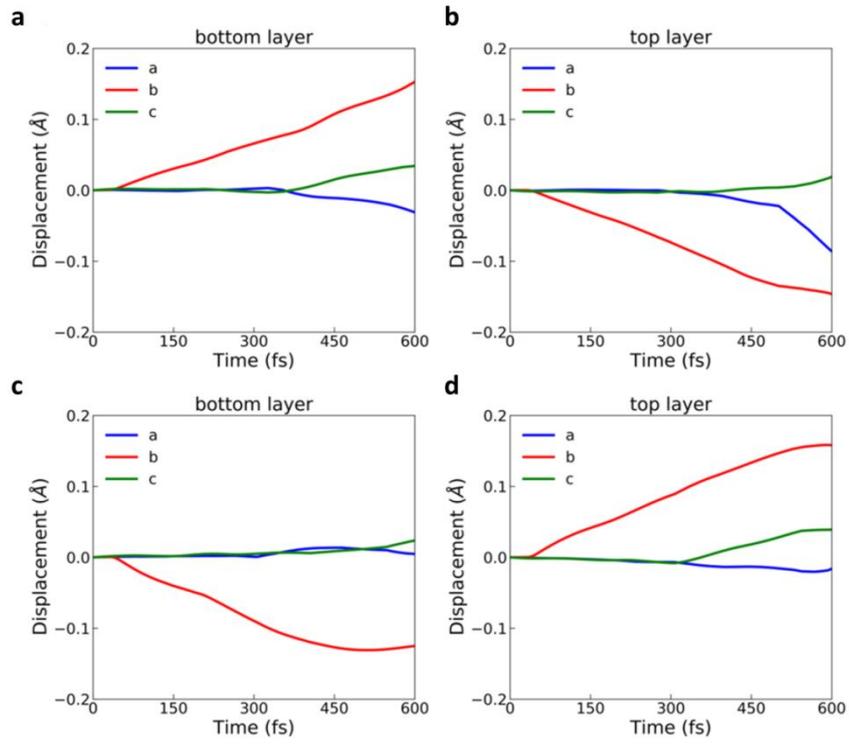

**Fig. S3| Interlayer displacements along the *a*-axis and *c*-axis with the photon energy of 0.6 eV. a,b,** The averaged atomic displacement of one layer along the *a*-axis (blue lines), *b*-axis (red lines) and *c*-axis (green lines) under LP-a excitations. **c,d,** are analogous to **a,b,** but under LP-b excitation.

**Supplementary Note 4: Suitable photon energy that can induce polarization-anisotropic shear mode**

Several photon energies are tested to confirm that the polarization-anisotropic response is determined by the elaborate electronic excitations. We found that when the photon energy is in the range of 0.5~0.8 eV, similar selective excitation of WSMs will emerge, leading to the polarization dependent interlayer displacement. Figure S4 show the results with three photon energies. Note that under smaller photon energy, (e.g., 0.5 eV), the interlayer displacement is larger, which might be ascribed to the fact that the transitions are closer to the Weyl nodes.

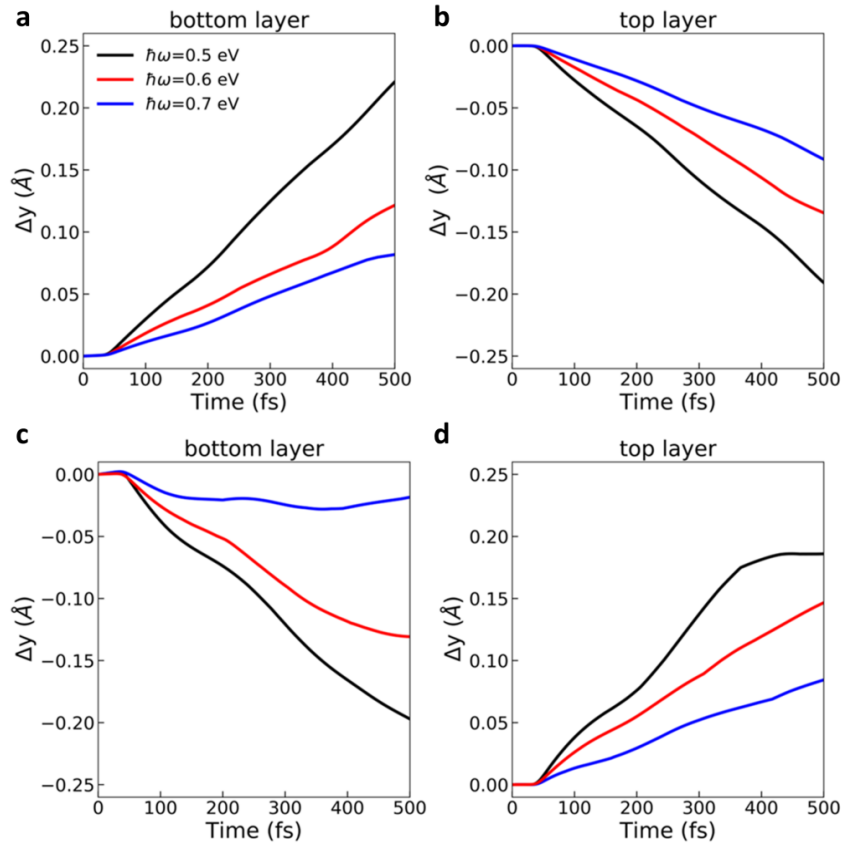

**Fig. S4| Photon-energy dependence of interlayer shear displacement.** **a**,**b**, Time-evolutions of interlayer shear displacement of the bottom (**a**) and the top (**b**) layer under LP-a excitation. **c,d,** are analogous to **a,b**, but under LP-b excitation. The laser duration and intensity are same as those shown in Fig. S1a for all the light pulses.

**Supplementary Note 5: Non-centrosymmetric degree of WTe$_2$ during laser illumination**

Figure S5a and S5b show the lattice structure of two WTe$_2$ phases, i.e., non-centrosymmetric $T_d$ phase and centrosymmetric 1T′(*) phase. Both two phases have an orthorhombic unit cell and can transform each other via interlayer shear displacement[37]. To characterize the non-centrosymmetric degree (NCD) of WTe$_2$ under laser illumination, a structure factor $d$ is introduced, which represents the structural difference between the two phases. The initial value of $d$ is 0.405 Å, meaning that if the bottom-layer displacement with respect to the top-layer is 0.405 Å and the two layers move towards each other, the $T_d$ phase will transform to 1T′(*) phase. The decrease of $d$ leading to the restoring of inversion symmetry, whereas the increase of $d$ indicates that the non-centrosymmetric order is further enhanced. Figure S5(c) shows the evolution of $d$ under the linearly polarized laser pulses. The decrease of $d$ under LP-a excitation can be compared with the time-evolution of experimental detected SHG intensity (Fig. 4(e) in Ref. [37]). In the main text, the normalized $d$ and SHG intensity are displayed in Fig. 5(c).

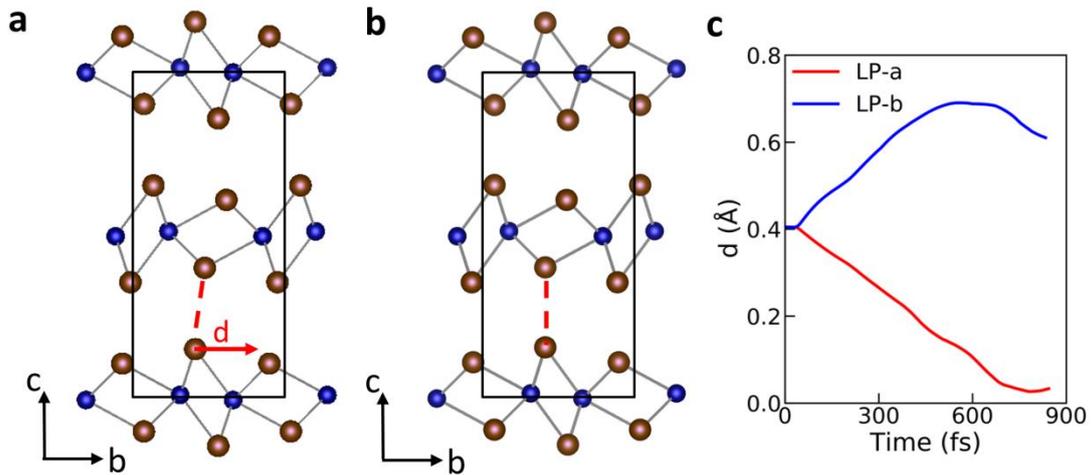

**Fig. S5| Non-centrosymmetric degree of WTe$_2$ during laser illumination. a,b,** Lattice structure of $T_d$ (a) and 1T′(*) (b) phases of WTe$_2$, structure factor $d$ is used to represent the interlayer difference between the two phases. **c,** Time-dependent $d$ under LP-a and LP-b excitations.